\documentclass[a4paper,11pt]{article}%
\usepackage{amssymb,amsmath,amsfonts}
\usepackage{graphicx}%
\usepackage{amsmath}%
\usepackage{amsfonts}%
\usepackage{amssymb}
\providecommand{\U}[1]{\protect\rule{.1in}{.1in}}
\newtheorem{theorem}{Theorem}

\newtheorem{corollary}[theorem]{Corollary}

\newtheorem{definition}[theorem]{Definition}

\newtheorem{proposition}[theorem]{Proposition}

\begin{document}

\title{How to Observe Quantum Fields and Recover \\Them from Observational Data?\\-- Takesaki Duality as a Micro-Macro Duality --}
\author{Izumi OJIMA and Mitsuharu TAKEORI
\and Research Institute for Mathematical Sciences, Kyoto University
\and Kyoto 606-8502, Japan}
\date{}
\maketitle

\begin{abstract}
After the mathematical notion of \textquotedblleft Micro-Macro
Duality\textquotedblright\ for understanding mutual relations between
microsopic quantum systems (Micro) and their macroscopic manifestations
(Macro) is explained on the basis of the notion of sectors and order
parameters, a general mathematical scheme is proposed for detecting the
state-structure inside of a sector through measurement processes of a maximal
abelian subalgebra of the algebra of observables. For this purpose, the
Kac-Takesaki operators controlling group duality play essential roles, which
naturally leads to the composite system of the observed system and the
measuring system formulated by a crossed product. This construction of
composite systems will be shown to make it possible for the Micro to be
reconstructed from its observational data as Macro in the light of the
Takesaki duality for crossed products.

\end{abstract}

\section{Quantum-Classical Correspondence and Micro-Macro Duality}

The essence of \textquotedblleft quantum-classical
correspondence\textquotedblright\ (q-c correspondence, for short) is usually
understood in such an intuitive way that \textit{macroscopic classical}
objects arise from a \textit{microscopic quantum} system as
\textit{condensates of infinite quanta} in the latter. Aiming at a
satisfactory understanding of mutual relations among different hierarchical
levels in the physical world, we try here to provide this heuristic idea with
a mathematically sound formulation on the basis of what we call
\textquotedblleft Micro-Macro duality\textquotedblright\ \cite{Oji05}; this is
the mathematical notion of \textit{duality} (or,\textit{\ categorical
adjunction }in more general situations), which allows us to connect
microscopic and macroscopic levels in the physical nature in
\textit{bi-directional} ways from Micro to Macro and vice versa. To be
precise, the contrasts of [Micro vs. Macro] (according to length scales) and
of [Quantum vs. Classical] (due to the essential differences in their
structures) are to certain extent independent of each other, as exemplified by
the presence of such interesting phenomena as \textquotedblleft macroscopic
quantum effects\textquotedblright\ owing to the absence of an intrinsic length
scale to separate quantum and classical domains. Since this kind of
mixtures\ can usually be taken as `exceptional', we restrict, for simplicity,
our consideration here to such generic situations that processes taking place
at microscopic levels are of quantum nature to be described by non-commutative
quantities and that the macroscopic levels are described in the standard
framework of classical physics in terms of commutative variables, unless the
considerations on the above point become crucial.

First we note that, in formulating a physical theory, we need the following
four basic ingredients, algebra of physical variables, its states and 
representations, their dynamical changes and a classifying space to classify,
describe and interpret the obtained theoretical and experimental results,
among which the algebra and its representations are mutually dual. When we try
to provide collected experimental results with a physical interpretation, the
most relevant points of the discussion starts from the problem to identify the
states responsible for the phenomena under consideration. On the premise of
the parallelisms among micro / quantum / non-commutative and macro / classical
/ commutative, respectively, the essential contents of {q-c} correspondence
can be examined in the following steps and forms:

1) Superselection \textit{sectors} and \textit{intersectorial structures}
described by order parameters: the first major gap between the microscopic
levels described by non-commutative algebras of physical variables and the
macroscopic ones by commutative algebras can be concisely formulated and
understood in terms of the notion of a (\textit{superselection}%
)\textit{\ sector structure }consisting of a family of \textit{sectors }(or
\textit{pure phases}) described mathematically by factor states and
representations: the algebra of observables is represented within a sector by
isomorphic von Neumann algebras with trivial centres and representations
corresponding to different sectors are mutually disjoint. The totality of
sectors (relevant to a given specific physical situation) constitutes
physically a \textit{mixed phase} involving both classical and quantum
aspects. Sectors or pure phases are faithfully parametrized by the spectrum of
the centre of a relevant representation of the C*-algebra of microscopic
quantum observables describing a physical system under consideration.
Physically speaking, operators belonging to the centre are mutually
commutative classical observables which can be interpreted as
\textit{macroscopic order parameters}. In this way, the intersectorial
structure describes the coexistence of and the gap between
quantum(=intrasectorial) and classical(=intersectorial) aspects.

2) \textit{Intrasectorial} quantum structures and measurement processes: it is
evident, however, that we cannot attain a satisfactory description of a given
quantum system unless we succeed in analyzing and describing the intrinsic
quantum structures \textit{within} a given sector, not only theoretically but
also operationally (up to the resolution limits imposed by quantum theory
itself). The detection of these invisible microscopic quantum aspects
necessarily involves the problem of quantum measurements. In the usual
discussions in quantum mechanics, a maximal abelian subalgebra (MASA, for
short)\ plays canonical roles in specifying a quantum state according to
measured data, in place of the centre trivialized by Stone-von Neumann theorem
of the uniqueness (up to unitary equivalence) of irreducible representations
of CCR algebras with finite degrees of freedom. As seen below, the notion of
MASA $\mathcal{A}$ plays central roles also in our context, whereas it need be
reformulated, in such a quantum system as quantum fields with infinite degrees
of freedom, whose algebra $\mathcal{M}$ of observables may have
representations of non-type I. The present formulation will be seen also to
determine the precise form of the coupling between the object system and the
apparatus required for implementing a measurement process, on the basis of
which the central notion of \textit{instrument} can be concisely formulated.

3) Inverse problem to \textit{reconstruct the algebra} of Micro system from
the observational Macro data: the roles played by the above MASA and by its
measured data suggest the possibility for us to reconstruct a microscopic
non-commutative algebra $\mathcal{M}$ from the macroscopic information
$Spec(\mathcal{A})$ obtained by measuring the MASA $\mathcal{A}$, in parallel
with the structure theory of semisimple Lie algebras based on their root
systems corresponding to chosen Cartan subalgebras. In fact, we show in Sec.5
that this analogy precisely works by means of the Takesaki duality applied to
the crossed product $\mathcal{M}\rtimes_{\alpha}\mathcal{U}$ arising from the
above coupling between the object system and the apparatus, where
$\mathcal{U}$ is a locally compact group acting on $\mathcal{A}$ and
generating it: $\mathcal{A}=\mathcal{U}^{\prime\prime}$ (in combination with a
modest technical assumption of \textquotedblleft
semi-duality\textquotedblright). This observation is conceptually very
important as the supporting evidence for the above-mentioned
\textit{bi-directionality} expected naturally in the notion of q-c
correspondence. Here, the notion of \textit{co-action }$\hat{\alpha}$ of
$\mathcal{\hat{U}}$ on $\mathcal{M}\rtimes_{\alpha}\mathcal{U}$ plays a
crucial role to reproduce $\mathcal{M}=(\mathcal{M}\rtimes_{\alpha}%
\mathcal{U})\rtimes_{\hat{\alpha}}\mathcal{\hat{U}}$ from $\mathcal{M}%
\rtimes_{\alpha}\mathcal{U}$, according to which the von Neumann type of
$\mathcal{M}$ is determined by the abelian dynamical system $[\mathfrak{Z}%
(\mathcal{M}\rtimes_{\alpha}\mathcal{U})=\mathcal{A}]\underset{\hat{\alpha}%
}{\curvearrowleft}\mathcal{\hat{U}}$ on the classifying space
$Spec(\mathcal{A})$ of the intrasectorial structure. On this last point, the
presence of certain subtle points is exhibited in relation with the essential
features of quantum systems with infinite degrees of freedom in the following form:

4) Roles of \textit{intrinsic dynamics} responsible for the recovery of
\textit{non-type I} algebras: when the algebra of observables is represented
in a Hilbert space as a von Neumann algebra of non-type I (like the typical
case\textit{\ }with local subalgebras of \textit{type III} in relativistic
quantum field theory), a state vector within a sector cannot uniquely be
specified by means of quantum observables for \textit{lack of minimal
projections}. In view of the above von Neumann-type classification, the
compatibility between the above reconstruction of $\mathcal{M}$ in 3) and its
non-trivial type forces the action $\alpha$ of $\mathcal{U}$ on $\mathcal{M}$
to deviate from the adjoint form $Ad_{u}(X)=uXu^{-1}$, which invalidates the
usual approximation adopted in most discussions of measurements to neglect the
\textit{intrinsic dynamics }of the object system closing up the coupling
between the system and the apparatus. In this case, therefore, we need the
information not only about states of the system but also its \textit{intrinsic
dynamics}, which can operationally be provided by the measurement of
energy-momentum tensor. In this way, all the basic ingredients, $\mathcal{M}$,
states on $\mathcal{M}$, the dynamics $\alpha$ and the classifying space
$Spec(\mathcal{A})$ of the intrasectorial structure exhibit themselves in the
discussion and to be determined operationally.

\subsection{Q-C correspondence (I): Sectors \& centre = order parameters}

At the level of \textit{sectors}, quantum and classical aspects can be
separated in a clear-cut way by means of order parameters to specify a sector.
To see this, we first recall the standard notion of \textit{quasi-equivalence}
$\pi_{1}\thickapprox\pi_{2}$ \cite{Dix} of representations $\pi_{1},\pi_{2}$
of an abstract C*-algebra $\mathfrak{A}$ describing the observables of a given
microscopic quantum system: taken as unitary equivalence \textit{up to
multiplicity}, this notion can be reformulated into many equivalent forms such
as the isomorphism of von Neumann algebras associated with representations:
\[
\pi_{1}\thickapprox\pi_{2}\Longleftrightarrow\pi_{1}(\mathfrak{A}%
)^{\prime\prime}\simeq\pi_{2}(\mathfrak{A})^{\prime\prime}\Longleftrightarrow
c(\pi_{1})=c(\pi_{2}),
\]
where $c(\pi)$ denotes the \textit{central support} of a representation $\pi$.
In the universal representation \cite{Dix} of $\mathfrak{A}$, $(\pi
_{u}:=\underset{{\omega}\in E_{\mathfrak{A}}}{\oplus}\pi_{\omega}%
,\mathfrak{H}_{u}:=\underset{{\omega}\in E_{\mathfrak{A}}}{\oplus}%
\mathfrak{H}_{\omega})$, $\pi_{u}(\mathfrak{A})^{\prime\prime}\simeq
\mathfrak{A}^{\ast\ast}=:\mathfrak{A}^{\prime\prime}$, consisting of all the
GNS representations $(\pi_{\omega},\mathfrak{H}_{\omega},\Omega_{\omega})$ for
states $\omega\in E_{\mathfrak{A}}$(: state space of $\mathfrak{A}$), the
central support $c(\pi)$ of $(\pi,\mathfrak{H}_{\pi}=P_{\pi}\mathfrak{H}_{u})$
with support projection $P_{\pi}\in\pi_{u}(\mathfrak{A})^{\prime}$ can be
defined as the smallest projection in the centre $\mathfrak{Z}(\mathfrak{A}%
^{\prime\prime}):=\mathfrak{A}^{\prime\prime}\cap\pi_{u}(\mathfrak{A}%
)^{\prime}$ to pick up all the representations quasi-equivalent to $\pi$:
$c(\pi)=$ projection onto $\overline{\pi(\mathfrak{A})^{\prime}\mathfrak{H}%
_{\pi}}\subset\mathfrak{H}_{u}$. On this basis, we introduce a basic scheme
for\textbf{\ }q-c correspondence in terms of sectors and order parameters: the
Gel'fand spectrum $Spec(\mathfrak{Z}(\mathfrak{A}^{\prime\prime}))$ of
$\mathfrak{Z}(\mathfrak{A}^{\prime\prime})$ arising from the \textquotedblleft
simultaneous diagonalization\textquotedblright\ of the commutative algebra
$\mathfrak{Z}(\mathfrak{A}^{\prime\prime})$ can be identified with the factor
spectrum $\overset{\frown}{\mathfrak{A}}$ \cite{Dix} of $\mathfrak{A}$:%
\[
Spec(\mathfrak{Z}(\mathfrak{A}^{\prime\prime}))\simeq\overset{\frown
}{\mathfrak{A}}:=F_{\mathfrak{A}}/\thickapprox:\mathbf{factor\ spectrum},
\]
defined by all the quasi-equivalence classes of \textit{factor} states
$\omega\in F_{\mathfrak{A}}$(: set of all factor states of $\mathfrak{A}$)
with trivial centres $\mathfrak{Z}(\pi_{\omega}(\mathfrak{A})^{\prime\prime
})=\pi_{\omega}(\mathfrak{A})^{\prime\prime}\cap\pi_{\omega}(\mathfrak{A}%
)^{\prime}=\mathbb{C}\mathbf{1}_{\mathfrak{H}_{\omega}}$ in the GNS
representations $(\pi_{\omega},\mathfrak{H}_{\omega})$.

\begin{definition}
A \textbf{sector (}or, physically speaking,\textbf{\ pure phase}) of
observable algebra $\mathfrak{A}$ is defined by a \textbf{quasi-equivalence
class of factor states} of $\mathfrak{A}$.
\end{definition}

In view of the commutativity of $\mathfrak{Z}(\mathfrak{A}^{\prime\prime})$
and of the role of its spectrum, we can regard \cite{Unif03}

\begin{itemize}
\item $\mathfrak{Z}(\mathfrak{A}^{\prime\prime})$ as the algebra of
\textbf{macroscopic order parameters} to specify sectors, and

\item $Spec(\mathfrak{Z}(\mathfrak{A}^{\prime\prime}))\simeq$ $\overset
{\frown}{\mathfrak{A}}$ as the \textbf{classifying space of sectors} to
distinguish among different sectors.
\end{itemize}

\noindent Then the dual map
\[
\text{\textbf{Micro: \ \ }}\mathfrak{A}^{\ast}\supset E_{\mathfrak{A}%
}\twoheadrightarrow Prob(\overset{\frown}{\mathfrak{A}})\subset L^{\infty
}(\overset{\frown}{\mathfrak{A}})^{\ast}\text{ \ : \textbf{Macro},}%
\]
of the embedding $\mathfrak{Z}(\mathfrak{A}^{\prime\prime})\simeq L^{\infty
}(\overset{\frown}{\mathfrak{A}})\hookrightarrow\mathfrak{A}^{\prime\prime}$
can be interpreted as a \textit{universal q(uantum)}$\rightarrow
$\textit{c(lassical) channel} which transforms a microscopic quantum state
$\phi\in E_{\mathfrak{A}}$ into a macroscopic classical state $\mu_{\phi}\in
Prob(\overset{\frown}{\mathfrak{A}})$ \cite{Unif03}:
\[
E_{\mathfrak{A}}\ni\phi\longmapsto\mu_{\phi}=\phi^{\prime\prime}%
\upharpoonright_{\mathfrak{Z}(\mathfrak{A}^{\prime\prime})}\in E_{\mathfrak{Z}%
(\mathfrak{A}^{\prime\prime})}=M^{1}(Spec(\mathfrak{Z}(\mathfrak{A}%
^{\prime\prime})))=Prob(\overset{\frown}{\mathfrak{A}})\,.
\]
$\mu_{\phi}$ is the probability distribution of sectors contained in a
mixed-phase state $\phi$ of $\mathfrak{A}$ in a quantum-classical composite
system,
\[
\overset{\frown}{\mathfrak{A}}\supset\Delta\longmapsto\phi^{\prime\prime}%
(\chi_{\Delta})=\mu_{\phi}(\Delta)=Prob(\text{sector}\in\Delta\text{ }|\text{
}\phi),
\]
wherer $\phi^{\prime\prime}$ denotes the normal extension of $\phi\in
E_{\mathfrak{A}}$ to $\mathfrak{A}^{\prime\prime}$. While it tells us as to
which sectors appear in $\phi$, it cannot specify as to which representative
factor state appears within each sector component of $\phi$. In other words,
our vocabulary at this level of resolution consists of words to indicate a
representation of $\mathfrak{A}$ as a whole which cannot pinpoint a specific
state belonging to it.

\subsection{Q-C correspondence: (II) Inside of sectors and maximal abelian
subalgebra}

To detect operationally the \textit{intrasectorial structures} inside of a
sector $\omega$ described by a factor representation $(\pi_{\omega
},\mathfrak{H}_{\omega},\Omega_{\omega})$, we need to choose a \textit{maximal
abelian subalgebra} (MASA) $\mathcal{A}$ of a factor algebra $\mathcal{M}%
:=\pi_{\omega}(\mathfrak{A})^{\prime\prime}$, characterized by the condition
$\mathcal{A}^{\prime}\cap\mathcal{M}=\mathcal{A}\cong L^{\infty}%
(Spec(\mathcal{A}))$ \cite{DixVN}. Note that, if we adopt the usual definition
of MASA, $\mathcal{A}^{\prime}=\mathcal{A}$, found in many discussions on
quantum-mechanical systems with finite degrees of freedom, the relation
$\mathcal{A}^{\prime}=\mathcal{A}\subset\mathcal{M}$ implies $\mathcal{M}%
^{\prime}\subset\mathcal{A}^{\prime}=\mathcal{A}\subset\mathcal{M}$, and
hence, $\mathcal{M}^{\prime}=\mathcal{M}^{\prime}\cap\mathcal{M}%
=\mathfrak{Z}(\mathcal{M})$ is of type I, which does not fit to the general
context of infinite systems involving algebras of non-type I. Since a tensor
product $\mathcal{M}\otimes\mathcal{A}$ (acting on the Hilbert-space tensor
product $\mathfrak{H}_{\omega}\otimes L^{2}(Spec(\mathcal{A}))$) has a centre
given by
\[
\mathfrak{Z}(\mathcal{M}\otimes\mathcal{A})=\mathfrak{Z}(\mathcal{M}%
)\otimes\mathcal{A}=\mathbf{1}\otimes L^{\infty}(Spec(\mathcal{A})),
\]
we see that the spectrum $Spec(\mathcal{A})$ of a MASA $\mathcal{A}$ to be
measured can be understood as parametrizing a \textit{conditional sector
structure} of the coupled system of the object system $\mathcal{M}$ and
$\mathcal{A}$, the latter of which can be identified with the measuring
apparatus $\mathcal{A}$ in the simplified version \cite{Unif03} of Ozawa's
measurement scheme \cite{Oza}. This picture of conditional sector structure is
consistent with the physical essence of a measurement process as
\textquotedblleft classicalization\textquotedblright\ of some restricted
aspects $\mathcal{A}$($\subset\mathcal{M}$) of a quantum system, conditional
on the coupling $\mathcal{M}\otimes\mathcal{A}$ of $\mathcal{M}$ with the
apparatus identified with $\mathcal{A}$.

In addition to the choice of relevant algebras of observables, the essential
point in the mathematical description of a measurement process is to find a
\textit{coupling term} between algebras $\mathcal{M}$ and $\mathcal{A}$ of
observables of the object system and of the measuring apparatus in such a way
that a microscopic quantum state of $\mathcal{M}$ can be determined by knowing
the macroscopic data of the pointer positions on $Spec(\mathcal{A})$ of the
measuring apparatus. To solve this problem we note that the algebra
$\mathcal{A}$ is generated by its unitary elements which constitute an abelian
unitary group $\mathcal{U}(\mathcal{A})$. As an infinite-dimensional group,
$\mathcal{U}(\mathcal{A})$ is, in general, not ensured to have an invariant
Haar measure. In the physically meaningful situations where observables are
represented in \textit{separable} Hilbert spaces, however, $\mathcal{A}$ as a
commutative von Neumann algebra can be shown to be generated by a single
element $A_{0}=A_{0}^{\ast}$: $\mathcal{A}=\{A_{0}\}^{\prime\prime}$
\cite{TakI}. This allows us to focus upon a one-parameter subgroup
$\{\exp(itA_{0});t\in\mathbb{R}\}$ of $\mathcal{U}(\mathcal{A})$ generating
$\mathcal{A}$ and equipped with an invariant Haar measure. In concrete
situations (where what is most relevant is as to which quantities are actually
measured), the existence of a \textit{single} generator valid at the level of
von Neumann algebras may sound too idealistic, but this point can easily be
remedied by relaxing it to a finite number of mutually commuting generators
consistently with the existence of a Haar measure. Thus, we treat in what
follows an abelian (Lie) group $\mathcal{U}$ equipped with a Haar measure $du$
which generates the MASA $\mathcal{A}$:
\[
\mathcal{U}\subset\mathcal{U}(\mathcal{A}),\mathcal{A}=\mathcal{U}%
^{\prime\prime}.
\]
Rewriting the condition $\mathcal{A}=\mathcal{A}^{\prime}\cap\mathcal{M}$ for
$\mathcal{A}$ to be a MASA of $\mathcal{M}$ into such a form as
\[
\mathcal{A}=\mathcal{M}\cap\mathcal{A}^{\prime}=\mathcal{M}\cap\mathcal{U}%
^{\prime}=\mathcal{M}^{\alpha(\mathcal{U})},
\]
we see that $\mathcal{A}$ is the fixed-point subalgebra of the adjoint action
$\alpha_{u}:=Ad(u):\mathcal{M}\ni X\longmapsto uXu^{\ast}$ of $\mathcal{U}$ on
$\mathcal{M}$ \cite{Oji05}. From this viewpoint, the relevance of the group
duality and of the Galois extension can naturally be expected. On the basis of
a formulation with a \textit{Kac-Takesaki operator} \cite{Tak69, NakTak}
(\textit{K-T operator}, for short) or a \textit{multiplicative unitary}
\cite{BaajSkan}, the universal essence of the problem can be understood in the
following form.

\section{Measurement Coupling and Instrument}

In the context of a \textit{Hopf-von Neumann algebra} $M(\subset
B(\mathfrak{H}))$ \cite{EnockSch} equipped with a Haar weight, a K-T operator
$V\in\mathcal{U}((M\otimes M_{\ast})^{-})\subset\mathcal{U}(\mathfrak{H}%
\otimes\mathfrak{H})$ is defined as the unitary implementer of its coproduct
$\Gamma:M\rightarrow M\otimes M$ in the sense of\ $\Gamma(x)=V^{\ast
}(\mathbf{1}\otimes x)V$. Corresponding to the co-associativity of $\Gamma$,
the K-T operator $V$ is characterized by the pentagonal relation,
$V_{12}V_{13}V_{23}=V_{23}V_{12}$, on $\mathfrak{H}\otimes\mathfrak{H}%
\otimes\mathfrak{H}$, where subscripts $i,j$ of $V_{ij}$ indicate the places
in $\mathfrak{H}\otimes\mathfrak{H}\otimes\mathfrak{H}$ on which the operator
$V$ acts. It plays most fundamental roles as an intertwiner, $V(\lambda
\otimes\iota)=(\lambda\otimes\lambda)V$, due to the quasi-equivalence among
tensor powers of the regular representation $\lambda:M_{\ast}\ni
\omega\longmapsto\lambda(\omega):=(i\otimes\omega)(V)\in\hat{M}$ given through
a generalized Fourier transform, $\lambda(\omega_{1}\ast\omega_{2}%
)=\lambda(\omega_{1})\lambda(\omega_{2})$, of the convolution algebra
$M_{\ast}$ with $\omega_{1}\ast\omega_{2}:=\omega_{1}\otimes\omega_{2}%
\circ\Gamma$. On these bases, a generalization of \textit{group duality} can
be formulated for Kac algebras \cite{EnockSch}. In the case of $M=L^{\infty
}(G,dg)$ with a locally compact group $G$ equipped with a (left-invariant)
Haar measure $dg$, the K-T operator $V$ is given on $L^{2}(G\times G)$ by%
\[
(V\xi)(s,t):=\xi(s,s^{-1}t)\text{ \ \ \ for }\xi\in L^{2}(G\times G),s,t\in
G,
\]
or symbolically, $V|s,t\rangle=|s,st\rangle$, in the Dirac-type notation.

To apply this machinery to our discussion involving the MASA $\mathcal{A}$, we
first recall the notion of the group dual $\hat{G}$ of a group $G$ defined by
the set of equivalence classes of irreducible unitary representations of $G$.
For our abelian group $\mathcal{U}$, its group dual $\widehat{\mathcal{U}}$
consists of the characters $\gamma$ of $\mathcal{U}$: $\gamma(u_{1}%
u_{2})=\gamma(u_{1})\gamma(u_{2}),$ $\gamma(e)=1$ ($u_{1}$, $u_{2}%
\in\mathcal{U}$). Identifying the above $M$ with $L^{\infty}(\widehat
{\mathcal{U}})=\lambda(\mathcal{U})^{\prime\prime}$, we consider the K-T
operator $V\in L^{\infty}(\widehat{\mathcal{U}})\otimes\lambda(\widehat
{\mathcal{U}})^{\prime\prime}=L^{\infty}(\widehat{\mathcal{U}}\times
\mathcal{U})$ associated with $\widehat{\mathcal{U}}$ taken as the above $G$:
\[
(V\xi)(\gamma_{1},\gamma_{2}):=\xi(\gamma_{1},\gamma_{1}^{-1}\gamma_{2})\text{
\ \ \ for }\xi\in L^{2}(\widehat{\mathcal{U}}\times\widehat{\mathcal{U}%
}),\gamma_{1},\gamma_{2}\in\widehat{\mathcal{U}},
\]
which satisfies the pentagonal relation, $V_{12}V_{13}V_{23}=V_{23}V_{12}$,
and the intertwining relation $V(\lambda_{\gamma}\otimes I)=(\lambda_{\gamma
}\otimes\lambda_{\gamma})V$ ($\gamma\in\widehat{\mathcal{U}}$) for the regular
representation $\widehat{\mathcal{U}}\ni\gamma\longmapsto\lambda_{\gamma}%
\in\mathcal{U}(L^{2}(\widehat{\mathcal{U}}\mathcal{))}$. We note here the
following implications of the inclusion relations $\mathcal{U}\subset
\mathcal{A}\subset\mathcal{M}$:

i) Any character $\chi\in Spec(\mathcal{A})$ of the MASA\ $\mathcal{A}$
defined as an \textit{algebraic homomorphism} $\chi:\mathcal{A}\rightarrow
\mathbb{C}$ is also a character $\chi\upharpoonright_{\mathcal{U}}\in
\widehat{\mathcal{U}}$ of the abelian unitary group $\mathcal{U}$ as a
\textit{group homomorphism} $\chi\upharpoonright_{\mathcal{U}}:\mathcal{U}%
\rightarrow\mathbb{T}$ by the restriction to $\mathcal{U}$. This implies the
inclusion $Spec(\mathcal{A})\hookrightarrow\widehat{\mathcal{U}}$, by which we
identify $\chi\in Spec(\mathcal{A})$ and $\chi\upharpoonright_{\mathcal{U}}%
\in\widehat{\mathcal{U}}$. While physically measured quantities would be
points $\chi$ in $Spec(\mathcal{A})$ which, in general, has no intrinsic base
point, the identity character $\iota\in\widehat{\mathcal{U}} $,$~\iota
(u)\equiv1$~($\forall u\in\mathcal{U}$) present in $\widehat{\mathcal{U}}$ can
be physically distinguished by its important role as the \textbf{neutral
position }of measuring pointer. To be precise, when $\mathcal{U}$ is not
compact, there is no vector $|\iota\rangle$ corresponding to $\iota\in
\widehat{\mathcal{U}}$ in $L^{2}(\mathcal{U})$, which can, however, be
remedied by replacing $\langle\iota|\cdots|\iota\rangle$ with the
\textit{invariant mean} $m_{\mathcal{U}}$ meaningful for all such amenable
groups as the abelian group $\mathcal{U}$. The importance of this neutral
position remarked earlier by Ozawa has been overlooked in the usual approaches
for lack of the suitable place to accommodate it in an intrinsic way.

ii) The inclusion map $E:\mathcal{A}=L^{\infty}(Spec(\mathcal{A}%
))\hookrightarrow\mathcal{M}$ defines an $\mathcal{M}$-valued spectral measure
$dE$ on $Spec(\mathcal{A})$ by $E(\Delta)=E(\chi_{\Delta})$ for Borel sets
$\Delta\subset Spec(\mathcal{A})$, and its restriction to $\mathcal{U}$
induces a spectral decomposition of $\mathcal{U}$ (as an application of the
SNAG theorem):
\[
E(u)=\int_{\chi\in Spec(\mathcal{A})\subset\widehat{\mathcal{U}}}%
\overline{\chi(u)}dE(\chi)\text{ \ \ \ }(u\in\mathcal{U}).
\]
Then the group homomorphism $\mathcal{U}\ni u\longmapsto E(u)\in\mathcal{M}$
gives an $\mathcal{M}$-valued unitary representation $E$ of the group
$\mathcal{U}$ in a Hilbret space $\mathfrak{H}_{\mathcal{M}}$ of $\mathcal{M}$
with spectral support given by $Spec(\mathcal{A})$:%
\[
supp(E)=supp(dE)=Spec(\mathcal{A})(\subset\widehat{\mathcal{U}}),
\]
where we can take $\mathfrak{H}_{\mathcal{M}}\ $as$\ L^{2}(\mathcal{M})$ (a
non-commutative $L^{2}$-space of $\mathcal{M}$), the Hilbert space where
$\mathcal{M}$ is represented in its standard form so that any normal state
$\omega$ of $\mathcal{M}$ is expressed in a vectorial form: $\omega
(A)=\langle\xi_{\omega}|A\xi_{\omega}\rangle$. Corresponding to this
representation $E$ of $\mathcal{U}$, a representation $E_{\ast}(V)=\int
_{\chi\in Spec(\mathcal{A})}dE(\chi)\otimes\lambda_{\chi}$ of the K-T operator
$V$ on $L^{2}(\mathcal{M})\otimes L^{2}(\widehat{\mathcal{U}})$ is defined by
\begin{equation}
E_{\ast}(V)(\xi\otimes|\gamma\rangle)=\int_{\chi\in Spec(\mathcal{A})}%
dE(\chi)\xi\otimes|\chi\gamma\rangle,\text{ \ \ \ for }\gamma\in
\widehat{\mathcal{U}},\text{ \ \ \ }\xi\in L^{2}(\mathcal{M}), \label{key}%
\end{equation}
satisfying the modified pentagonal relation $E_{\ast}(V)_{12}E_{\ast}%
(V)_{13}V_{23}=V_{23}E_{\ast}(V)_{12}$.

iii) In view of the inclusion relations $E(u)=u\in\mathcal{U}\subset
\mathcal{A}\subset\mathcal{M}$, it may appear strange or pedantic to introduce
the map $E$ and to talk about it as a unitary \textit{representation}
$(E,L^{2}(\mathcal{M}))$ of $\mathcal{U}$ in $\mathfrak{H}_{\mathcal{M}}$. As
will be shown later, however, this is not the case, since it turns out to be
crucial to distinguish $\mathcal{U}$ itself as an \textquotedblleft%
\textit{abstract}\textquotedblright\ group from the \textit{represented}
unitary group $\mathcal{U}\subset\mathcal{M}$ embedded in $\mathcal{M}$.
First, we note that the group $\mathcal{U}$ has the \textit{regular
representation} $(\lambda,L^{2}(\mathcal{U},du))$ as its\ canonically defined
representation in the Hilbert space $L^{2}(\mathcal{U},du)$ with the Haar
measure $du$ of $\mathcal{U}$, which is isomorphic to the Hilbert space
$L^{2}(\widehat{\mathcal{U}},d\gamma)$ through the \textit{Fourier transform}
$\mathcal{F}$ from $\mathcal{U}$ to $\widehat{\mathcal{U}}$ as a unitary
transformation given by
\begin{equation}
(\mathcal{F}\xi)(\gamma)=\int_{\mathcal{U}}\overline{\gamma(u)}\xi(u)du,\text{
\ \ \ }(\mathcal{F}^{-1}\eta)(u)=\int_{\widehat{\mathcal{U}}}\gamma
(u)\eta(\gamma)d\gamma. \label{FT}%
\end{equation}
While$\ \mathcal{U}(\subset\mathcal{M})$ and $\lambda(\mathcal{U})(\subset
B(L^{2}(\mathcal{U})))$ are isomorphic \textit{as groups}, the corresponding
von Neumann algebras given by their weak closures are, in general, different,
$\mathcal{U}^{\prime\prime}=\mathcal{A}\hookrightarrow L^{\infty}%
(\widehat{\mathcal{U}})$, $Spec(\mathcal{A})\subset\widehat{\mathcal{U}}$,
owing to the difference in their representation Hilbert spaces: the former is
represented through the action of $\mathcal{U}$ on $\mathcal{M}$ in the
representation space $\mathfrak{H}_{\mathcal{M}}$ of $\mathcal{M}$, and the
latter in $L^{2}(\mathcal{U})\cong L^{2}(\widehat{\mathcal{U}})$. As will be
shown in Sec.4, the differences between $\mathcal{U}^{\prime\prime
}=\mathcal{A}$ and $L^{\infty}(\widehat{\mathcal{U}})=\lambda(\mathcal{U}%
)^{\prime\prime}$, or between $Spec(\mathcal{A})$ and $\widehat{\mathcal{U}}$,
determine the von Neumann type of $\mathcal{M}$, according to which
$\mathcal{A}\cong L^{\infty}(\widehat{\mathcal{U}})$, or equivalently,
$Spec(\mathcal{A})=\widehat{\mathcal{U}}$, holds if and only if $\mathcal{M}$
is of type I.

Now the important operational meaning of the equality (\ref{key}) and the role
of the neutral position $\iota$ can clearly be seen, especially if
$\widehat{\mathcal{U}}$ is a \textit{discrete} group which is equivalent to
the \textit{compactness} of the group $\mathcal{U}$: choosing $\chi=\iota$, we
have the equality, $E_{\ast}(V)(\xi_{\gamma}\otimes|\iota\rangle)=\xi_{\gamma
}\otimes|\gamma\rangle$\ ($\forall\gamma\in\widehat{\mathcal{U}}$,
$\xi_{\gamma}\in E(\gamma)\mathfrak{H}_{\mathcal{M}}$). With a generic state
$\xi=\sum_{\gamma\in\widehat{\mathcal{U}}}c_{\gamma}\xi_{\gamma}$ of
$\mathcal{M}$, an initial \textit{uncorrelated\ }state\textit{\ }$\xi
\otimes|\iota\rangle$ is transformed by $E_{\ast}(V)$ to a \textit{correlated}
one:%
\[
E_{\ast}(V)(\xi\otimes|\iota\rangle)=\sum_{\gamma\in\widehat{\mathcal{U}}%
}c_{\gamma}\xi_{\gamma}\otimes|\gamma\rangle.
\]
If $\mathcal{M}$ is not of type III equipped with a normal faithful
semi-finite (n.f.s., for short) trace, this establishes a \textit{one-to-one
}correspondence (\textquotedblleft perfect correlation\textquotedblright\ due
to Ozawa \cite{Ozawa03}) between a state $|\gamma\rangle$ of the measuring
probe system $\mathcal{A}$ specified by an observed value $\gamma\in
Spec(\mathcal{A})$ on the pointer and the corresponding unique state
$\xi_{\gamma}\in\mathcal{M}_{\gamma}$ of the microscopic system $\mathcal{M}$.
If $\mathcal{M}$ is of type III, $\dim(\mathcal{M}_{\gamma})\leq1$ is not
guaranteed for lack of a trace, and hence, the notion of perfect correlation
may fail to hold in such cases. Moreover, if we find some evidence for such a
kind of uncertainty as violating $\dim(\mathcal{M}_{\gamma})\leq1$, then it
implies that $\mathcal{M}$ should be of type III.

On these bases, we can define the notion of an\textbf{\ instrument}%
\textit{\textbf{\ }}$\mathfrak{I}$ as a (completely) positive operation-valued
measure to unify all the ingredients relevant to a measurement as follows:
\begin{align*}
\mathfrak{I}(\Delta|\omega_{\xi})(B)  &  {:=}(\omega_{\xi}\otimes
m_{\mathcal{U}})(E_{\ast}(V)^{\ast}(B\otimes\chi_{\Delta})E_{\ast}(V))\\
&  =(\langle\ \xi|\otimes\langle\iota|)E_{\ast}(V)^{\ast}(B\otimes\chi
_{\Delta})E_{\ast}(V)(|\xi\rangle\otimes|\iota\rangle).
\end{align*}
In the situation with a state $\omega_{\xi}=\langle\ \xi|\ (-)\xi\rangle$ of
$\mathcal{M}$ as an initial state of the system, the instrument describes
simultaneously the probability $p(\Delta|\omega_{\xi})=\mathfrak{I}%
(\Delta|\omega_{\xi})(\mathbf{1})$ for measured values of observables in
$\mathcal{A}$ to be found in a Borel set $\Delta$ and the final state
$\mathfrak{I}(\Delta|\omega_{\xi})/p(\Delta|\omega_{\xi})$ realized through
the detection of measured values \cite{Oza}. The merits of the present
formulation of instrument consist in such points that it is free from the
restriction on the types of von Neumann algebras and that it can be applied to
\textit{any} measurement, irrespective of whether repeatable or not, since any
drastic changes between initial and final states can be easily absorbed in the
system with \textit{infinite degrees of freedom}.

\section{Crossed Product $\mathcal{M}\rtimes_{\alpha}\mathcal{U}$ as Composite
System of System \&\ Apparatus}

Here we clarify the important meaning of the coupling $E_{\ast}(V)$ between
the system $\mathcal{M}$ to be observed and the measuring apparatus
corresponding to a MASA$\ \mathcal{A}$ of $\mathcal{M}$: its essential roles
in the whole measurement processes are closely related with the crossed
product $\mathcal{M}\rtimes_{\alpha}\mathcal{U}$ to describe the composite
system of $\mathcal{M}$ and $\mathcal{A}$ to be put in the context of
\textit{\textbf{Fourier-Galois duality}} and with the amplification processes
for the measured data $\chi\in Spec(\mathcal{A})\subset\widehat{\mathcal{U}}$
to take macroscopically visible forms emerging from the small changes at the
microscopic tip of the measuring apparatus caused by this coupling. For this
purpose, we consider the \textit{Fourier transform} of the K-T operator $V\in
L^{\infty}(\widehat{\mathcal{U}})\otimes\lambda(\widehat{\mathcal{U}}%
)^{\prime\prime}$ on $\widehat{\mathcal{U}}$, $(V\xi)(\gamma_{1},\gamma
_{2})=\xi(\gamma_{1},\gamma_{1}^{-1}\gamma_{2})$ ($\xi\in L^{2}(\widehat
{\mathcal{U}}\times\widehat{\mathcal{U}})$), given by
\begin{align*}
&  W:=(\mathcal{F}\otimes\mathcal{F})^{-1}V(\mathcal{F}\otimes\mathcal{F}),\\
&  (W\xi)(u_{1},u_{2}):=\xi(u_{2}u_{1},u_{2})\text{ \ \ \ for }\xi\in
L^{2}(\mathcal{U}\times\mathcal{U}),u_{1},u_{2}\in\mathcal{U}.
\end{align*}
This $W\in\lambda(\mathcal{U})^{\prime\prime}\otimes L^{\infty}(\mathcal{U})$
is seen also to be a K-T operator on $\mathcal{U}$ belonging to $\lambda
(\mathcal{U})^{\prime\prime}\otimes L^{\infty}(\mathcal{U})$ characterized by
the pentagonal and the intertwining relations:
\begin{align*}
W_{12}W_{13}W_{23}  &  =W_{23}W_{12},\\
W(\lambda_{u}\otimes\lambda_{u})  &  =(I\otimes\lambda_{u})W,\text{
\ \ \ }(u\in\mathcal{U}),
\end{align*}
for the regular representation $\lambda=\lambda^{\mathcal{U}}$ of
$\mathcal{U}$ on $L^{2}(\mathcal{U})$. Through the embedding map
$E:\mathcal{U}\hookrightarrow\mathcal{A}\hookrightarrow\mathcal{M}$, this K-T
operator $W$ is represented in $\mathcal{M}$ by $EW:=(E\otimes id)(W)\in
\mathcal{A}\otimes L^{\infty}(\mathcal{U})\subset\mathcal{M}\otimes L^{\infty
}(\mathcal{U}))$, which satisfies the modified version of pentagonal and
intertwining relations:
\begin{align*}
(EW)_{12}(EW)_{13}W_{23}  &  =W_{23}(EW)_{12},\\
EW(u\otimes\lambda_{u})  &  =(I\otimes\lambda_{u})EW.
\end{align*}
In view of the relation
\begin{equation}
\lbrack(EW)\hat{X}(EW^{\ast})](u)=u^{-1}\hat{X}(u)u=\alpha_{u}^{-1}(\hat
{X}(u))
\end{equation}
valid for $\hat{X}\in\mathcal{M}\otimes L^{\infty}(\mathcal{U})$, we define an
injective *-homomorphism $\pi_{\alpha}:\mathcal{M}\rightarrow L^{\infty
}(\mathcal{U},\mathcal{M})=\mathcal{M}\otimes L^{\infty}(\mathcal{U})$ by
\begin{align}
(\pi_{\alpha}(X)\xi)(u)  &  :=\alpha_{u}^{-1}(X)(\xi(u))=(u^{-1}Xu)(\xi(u))\\
&  \text{for }\xi\in L^{2}(\mathcal{M})\otimes L^{2}(\mathcal{U}%
),u\in\mathcal{U},\nonumber
\end{align}
which is implemented by $EW$:
\[
\pi_{\alpha}(X)=(EW)(X\otimes I)(EW)^{\ast}\ \ \text{\ for }X\in\mathcal{M}.
\]
According to \cite{NakTak}, the von Neumann algebra generated by $\pi_{\alpha
}(\mathcal{M})$ and $\mathbb{C}I\otimes\lambda(\mathcal{U})^{\prime\prime}$ is
just a crossed product $\mathcal{M}\rtimes_{\alpha}\mathcal{U}$:
\[
\mathcal{M}\rtimes_{\alpha}\mathcal{U}:=\pi_{\alpha}(\mathcal{M}%
)\vee(\mathbb{C}\otimes\lambda(\mathcal{U})^{\prime\prime}).
\]
This can also be viewed as (the weak-operator closure of) the image of the
convolution *-algebra $L^{1}(\mathcal{U},\mathcal{M})=\mathcal{M}\otimes
L^{1}(\mathcal{U})$ with the product\ structure given for $X,Y\in
L^{1}(\mathcal{U},\mathcal{M})$ by%
\begin{align*}
(X\ast Y)(u)  &  =\int_{\mathcal{U}}X(v)\alpha_{v}(Y(v^{-1}u))dv,\\
X^{\#}(u)  &  =\alpha_{u}(X(u^{-1}))^{\ast},
\end{align*}
under the operator-valued Fourier transform $\mathfrak{F}$:
\begin{align*}
&  \mathfrak{F}(X) =(Xdu\otimes id)(\sigma(EW)^{\ast}\sigma)=\int
_{\mathcal{U}}X(u)udu\text{ \ \ \ }\\
&  \text{ \qquad\qquad\ for }X \in L^{1}(\mathcal{U},\mathcal{M}%
)=\mathcal{M}\otimes L^{1}(\mathcal{U});\\
&  \mathfrak{F}(X\ast Y) =\mathfrak{F}(X)\mathfrak{F}(Y)\text{ \ and
}\mathfrak{F}(X^{\#})=\mathfrak{F}(X)^{\ast},
\end{align*}
where $\sigma$ is the flip operator interchanging tensor factors: $\sigma
(\xi\otimes\eta):=\eta\otimes\xi$. In this way, the crucial roles played by
the coupling $EW$ between the observed system $\mathcal{M}$ and the probe
system $\mathcal{A}$ can be seen in the the formation of their composite
system in the form of a crossed product $\mathcal{M}\rtimes_{\alpha
}\mathcal{U}$. In the process $\iota\rightarrow\alpha\rightarrow\iota$ of
switching-on and -off the coupling $\alpha$ starting from $\alpha=\iota$, the
structure of $\mathcal{M}\rtimes_{\alpha}\mathcal{U}$ will be seen to change
as $\mathcal{M}\otimes L^{\infty}(\widehat{\mathcal{U}})\rightarrow
\mathcal{M}\rtimes_{\alpha}\mathcal{U}\rightarrow\mathcal{M}\otimes L^{\infty
}(\widehat{\mathcal{U}})$.

\subsection{Physical meaning of crossed product and Takesaki duality}

The importance of the crossed product $\mathcal{M}\rtimes_{\alpha}\mathcal{U}$
can be seen in the relation with the Takesaki duality \cite{T1}:
\[
(\mathcal{M}\rtimes_{\alpha}\mathcal{U)}\rtimes_{\hat{\alpha}}\widehat
{\mathcal{U}}\simeq\mathcal{M}\otimes B(L^{2}(\mathcal{U}))\simeq\mathcal{M},
\]
where the last isomorphism $\mathcal{M}\simeq\mathcal{M}\otimes B(L^{2}(G))$
holds for any \textit{properly infinite} von Neumann algebras $\mathcal{M}$ as
applies to the present situation discussing a quantum system with infinite
degreees of freedom. Here $\hat{\alpha}$ is the dual co-action \cite{NakTak}
of $\mathcal{U}$ on $\mathcal{M}\rtimes_{\alpha}\mathcal{U}$ defined for
$Y\in\mathcal{M}\rtimes_{\alpha}\mathcal{U}$ by
\[
\pi_{\hat{\alpha}}(Y):=Ad(1\otimes\sigma W^{\ast}\sigma)(Y\otimes1),
\]
which reduces just to the action of the group dual $\widehat{\mathcal{U}}$ on
$\mathcal{M}\rtimes_{\alpha}\mathcal{U}$ in the case of abelian group
$\mathcal{U}$. In this context, a crossed product $\mathcal{M}\rtimes_{\alpha
}\mathcal{U}$ can be viewed as a kind of the non-commutative Fourier dual of
$\mathcal{M}$ whose precise knowledge enables us to \textbf{recover} the
original \textbf{algebra} $\mathcal{M}$ of the microscopic quantum system by
forming the \textit{second crossed product} with the dual action $\hat{\alpha
}$ by $\widehat{\mathcal{U}}$. Our original purpose of considering the
composite system $\mathcal{M}\rtimes_{\alpha}\mathcal{U}$ was to prepare a
measurement process just for analyzing the structure of \textit{states} within
a sector starting from the postulated knowledge of the algebra $\mathcal{M}$
of a Micro-system on the basis of the coupling term to yield experimental data
in $Spec(\mathcal{A})\subset\widehat{\mathcal{U}}$. The fullfilment of this
step, however, drives us into the next step in the opposite direction of
\textit{reconstructing} the \textit{original algebra} $\mathcal{M}$ from the
observational data on states. As a result, the essential idea of
\textit{Micro-Macro duality} \cite{Oji05} is implemented mathematically by the
duality of crossed products as an operator-algebraic extension of
Fourier-Galois duality: if the algebra $\mathcal{M}$ of the Micro-system is
known beforehand for one reason or another, this scheme can be used for
checking whether $\mathcal{M}$ is correctly chosen or not through the
comparison of the theoretical predicitions encoded in $\mathcal{M}%
\rtimes_{\alpha}\mathcal{U}$ and the actually observed data. On the other
hand, if $\mathcal{M}$ is \textit{unknown} (as is in the usual situations),
the latter data can serve for constructing $\mathcal{M}$ from which one should
rederive the observational data to ensure the consistency.

To proceed further, we add here a mathematical postulate called
\textit{semi-duality} \cite{NakTak} of the action $\alpha$ on $\mathcal{M}$,
which assumes the existence of such a unitary $v\in\mathcal{M}\otimes
\lambda(\mathcal{U})^{\prime\prime}$ that the condition $\overline{\alpha
}(v)=(v\otimes1)(1\otimes V^{\prime})$ holds with\ a K-T operator $V^{\prime}$
given by $(V^{\prime}\xi)(u_{1},u_{2})=\xi(u_{1}u_{2},u_{2})$ and
$\overline{\alpha}:=(\iota\otimes\sigma)\circ(\alpha\otimes\iota) $. From this
assumption follows the relation $(\mathcal{M}\otimes B(L^{2}(\mathcal{U}%
)))^{\tilde{\alpha}(\mathcal{U)}}=\mathcal{M}^{\alpha(\mathcal{U)}}\otimes
B(L^{2}(\mathcal{U}))$ (see \cite{NakTak}), which implies, in combination with
the relation $\mathcal{M}\rtimes_{\alpha}\mathcal{U}\simeq(\mathcal{M}\otimes
B(L^{2}(\mathcal{U})))^{\tilde{\alpha}(\mathcal{U)}}$ with $\tilde{\alpha
}=\alpha\otimes Ad\circ\lambda$ (valid for any crossed products with abelian
group actions), the following interesting structure for the crossed product:
\[
\mathcal{M}\rtimes_{\alpha}\mathcal{U}\simeq(\mathcal{M}\otimes B(L^{2}%
(\mathcal{U})))^{\tilde{\alpha}(\mathcal{U)}}=\mathcal{M}^{\alpha
(\mathcal{U)}}\otimes B(L^{2}(\mathcal{U}))=\mathcal{A}\otimes B(L^{2}%
(\mathcal{U})).
\]
We see that in this situation the Takesaki duality splits into two parts as follows:

\begin{theorem}
Let $\mathcal{M}$ and $\mathcal{A}=\mathcal{A}^{\prime}\cap\mathcal{M}$ be,
respectively, a properly infinite von Neumann algebra and its MASA generated
by a locally compact abelian unitary group $\mathcal{U\subset A}%
=\mathcal{U}^{\prime\prime}=\mathcal{M}^{\alpha(\mathcal{U)}}$. If the
semi-duality condition $\overline{\alpha}(v)=(v\otimes1)(1\otimes V^{\prime})$
holds for the action $\alpha$ of $\mathcal{U}$, then the Takesaki duality
\cite{T1} for $\mathcal{M}$ and $\mathcal{A}$, $(\mathcal{M}\rtimes_{\alpha
}\mathcal{U})\rtimes_{\hat{\alpha}}\widehat{\mathcal{U}}\simeq\mathcal{M}%
\otimes B(L^{\infty}(\mathcal{U}))\simeq\mathcal{M}$ and $(\mathcal{A}\otimes
B(L^{\infty}(\mathcal{U})))\rtimes_{\hat{\alpha}}\widehat{\mathcal{U}}%
)\rtimes_{\mu}\mathcal{U}\simeq\mathcal{A}\otimes B(L^{\infty}(\widehat
{\mathcal{U}}))$, can be decomposed into the following two mutually equivalent isomorphisms:

\begin{enumerate}
\item[i)] $\mathcal{M}\rtimes_{\alpha}\mathcal{U}\simeq\mathcal{A}\otimes
B(L^{\infty}(\mathcal{U}))$[: amplification process],

\item[ii)] $(\mathcal{A}\otimes B(L^{\infty}(\mathcal{U}))\rtimes_{\hat
{\alpha}}\widehat{\mathcal{U}}\simeq\mathcal{M}$[: reconstruction].
\end{enumerate}
\end{theorem}

By means of this, we can attain the following clear-cut mathematical
description of the physical situations relevant to measurement processes of
quantum dynamical systems with infinite degrees of freedom, which explains
both aspects at the same time, the amplification processes from invisible
Micro to visible Macro data and the recovery of invisible Micro from visible
Macro data.\noindent\ 

According to i), the composite system $\mathcal{M}\rtimes_{\alpha}\mathcal{U}$
of a Micro-quantum system $\mathcal{M}$ (of local fields, for instance) and of
a measuring apparatus coupled through an action $\alpha$ of the unitary group
$\mathcal{U}$ generating a MASA$\ \mathcal{U}^{\prime\prime}=\mathcal{A}%
=\mathcal{A}^{\prime}\cap\mathcal{M}$ can be decomposed into a classical
system with a commutative algebra $\mathcal{A}$ to be measured and a
quantum-mechanical one $B(L^{2}(\mathcal{U}))$ of CCR with finite degrees of
freedom. Arising from the Heisenberg group composed of two abelian groups
$\mathcal{U}$ and $\widehat{\mathcal{U}}$ in Fourier-Pontryagin duality, this
latter component will be seen to play physically interesting role as the
\textquotedblleft reservoir\textquotedblright\ in the relaxation processes of
\textit{amplification} to extract Macro from Micro; namely, the former half of
the Takesaki duality, $\mathcal{M}\rtimes_{\alpha}\mathcal{U}\simeq
\mathcal{A}\otimes B(L^{2}(\mathcal{U}))$, provides the mathematical basis for
the process to amplify the measured quantities in $\mathcal{A}$ into
macroscopically visible data at the expense of the dissipative damping effects
to suppress other irrelevant quantities. This picture is based upon the
following two points, one being the homotopical notion of strong Morita
equivalence and the other the quasi-equivalence of arbitrary tensor powers
$\lambda^{\otimes n}$ of regular representation $\lambda$ of $\mathcal{U}$.
The notion of strong Morita equivalence $\mathfrak{A}_{1}\thickapprox
\mathfrak{A}_{2}$ of algebras $\mathfrak{A}_{1}$, $\mathfrak{A}_{2}$ is
defined by the isomorphism $Rep_{\mathfrak{A}_{1}}\simeq Rep_{\mathfrak{A}%
_{1}}$ of their representation categories which is equivalent to the stability
$\mathfrak{A}_{1}\otimes\mathcal{K}\simeq\mathfrak{A}_{2}\otimes\mathcal{K}$
under tensoring the compact operator algebra $\mathcal{K}=\mathcal{K}%
(\mathfrak{H})$ \cite{Rieffel}. Physically this notion fits to the purpose of
ensuring the stability of the object system against noise perturbations coming
from its neglected surroundings. In the present context of focusing on the
internal structure of a sector $\omega$ of $\mathfrak{A}$, the
\textit{topological} form $\mathfrak{A}\thickapprox\mathfrak{A}\otimes
\mathcal{K}(L^{2}(\widehat{\mathcal{U}}))$ of Morita equivalence for the
C*-algebra $\mathfrak{A}$ of observables is converted into the
\textit{measure-theoretical} one $\mathcal{M}=\pi_{\omega}(\mathfrak{A}%
)^{\prime\prime}\simeq\mathcal{M}\otimes B(L^{2}(\widehat{\mathcal{U}}))$ as
the isomorphism of von Neumann algebras, which automatically holds for an
arbitrary \textit{properly infinite }von Neumann algebra $\mathcal{M}$
describing a quantum dynamical system with infinite degrees of freedom like
quantum fields. This allows us to interchange $\mathcal{M}$ and $\mathcal{M}%
\otimes B(L^{2}(\widehat{\mathcal{U}}))$ freely without any changes. On the
other hand, arbitrary tensor powers $(\lambda^{\otimes n},L^{2}(\mathcal{U}%
)^{\otimes n})$ ($n\in\mathbb{N}$) of regular representation $(\lambda
,L^{2}(\mathcal{U}))$ of $\mathcal{U}$ are quasi-equivalent via the K-T
operator $W$, and hence, the relation $\mathcal{M}\rtimes_{\alpha}%
\mathcal{U}\simeq\mathcal{A}\otimes B(L^{\infty}(\mathcal{U}))$ related with
$\tilde{\alpha}=\alpha\otimes Ad\circ\lambda$ can be extended to the situation
involving $\tilde{\alpha}^{(n)}=\alpha\otimes Ad\circ\lambda^{\otimes n}$.
This provides the mathematical support for the \textit{repeatability}
hypothesis of the measurement processes, which can be formulated consistently
in the framework of quantum stochastic processes. (Note that the distinction
between repeatable and non-repeatable ones disappears in the system with
infinite degrees of freedom.) In this context the notion of operator-valued
weights associated with the \textquotedblleft\textit{integrability}%
\textquotedblright\ \cite{NakTak} of an action following from the assumption
of semi-duality\ plays important roles. Along this line, the stochastic
processes developed on the tensor algebra generated by $(\lambda
,L^{2}(\mathcal{U}))$ are expected to provide a natural basis for the
processes to \textit{amplify} microscopic changes caused by the coupling
between the microscopic end (called a probe system) of measuring apparatus and
the observed microscopic system into the macroscopically visible motions of
the measuring pointer. Note at the same time, however, the sharp contrast
between the situations with $n=0$ and $n>1$, since the above isomorphism valid
for a properly infinite algebra $\mathcal{M}$ does not apply to the
MASA\ $\mathcal{A}$ which are commutative, and hence, \textit{not} properly
infinite. This will be seen also to be related with such complication that
uniqueness of MASA up to unitary conjugacy valid in a von Neumann algebra of
type I is not guaranteed in non-type I cases.

More interesting is the second isomorphism,
\[
\mathcal{M}\simeq\mathcal{M}\otimes B(L^{2}(\mathcal{U}))\simeq(\mathcal{M}%
\rtimes_{\alpha}\mathcal{U)}\rtimes_{\hat{\alpha}}\widehat{\mathcal{U}}%
\simeq(\mathcal{A}\otimes B(L^{2}(\mathcal{U})))\rtimes_{\hat{\alpha}}%
\widehat{\mathcal{U}}\text{, }%
\]
by which the \textit{invisible} microscopic algebra $\mathcal{M}$ of quantum
observables is \textit{recovered} from the information on the
\textit{macroscopically visible} MASA $\mathcal{A}$ with its measured valued
in the spectrum $Spec(\mathcal{A})$ together with that of the dual group
$\widehat{\mathcal{U}}$ of an abelian group $\mathcal{U}$ in $\mathcal{A}$ to
generate $\mathcal{A}=\mathcal{U}^{\prime\prime}$, both constituting the
quantum-mechanical CCR\ algebra $B(L^{2}(\mathcal{U}))$. As shown in the next
section, this is not merely a matter of interpretation but it actually
provides the crucial information on the von Neumann type classification of the
quantum algebra $\mathcal{M}$ on the basis of which the claimed
bi-directionality at the beginning is ensured.

\section{Reconstruction of Micro-Algebra $\mathcal{M}$ \& its
Type-Classification}

The main purpose here is to analyze the structure of the von Neumann factor
$\mathcal{M}$ describing a fixed sector from the viewpoint of ii),
$\mathcal{M}\simeq(\mathcal{A}\otimes B(L^{2}(\mathcal{U})))\rtimes
_{\hat{\alpha}}\widehat{\mathcal{U}}$, in the last section, in the systematic
use of the observable data provided by the measurement processes described by
i), $\mathcal{M}\rtimes_{\alpha}\mathcal{U}\simeq\mathcal{A}\otimes
B(L^{2}(\mathcal{U}))=:\mathcal{N}$. To achieve it in an effective way, we
need the description of the modular structure of $\mathcal{M}$ given as a
crossed product $\mathcal{M}=\mathcal{N}\rtimes_{\theta}G$ of the W*-dynamical
system $\mathcal{N}\underset{\theta}{\curvearrowleft}G$ in terms of its
component algebra $\mathcal{N}$ and the (co-)action $\theta:=\hat{\alpha}$ of
an locally compact abelian group $G=\widehat{\mathcal{U}}$.

\subsection{Dynamical systems and crossed products}

We first need some basic notions related to the W*-dynamical system
$\mathcal{N}\underset{\theta}{\curvearrowleft}G$, where the action
$\theta=\hat{\alpha}$ of $G$ on $\mathcal{N}$ is given in the form $\theta
_{g}=\beta_{g}\otimes AdU_{g}$ with $U_{g}$ a unitary representation of $G$ on
$\mathfrak{H}$. $\mathcal{N}$ can be identified with a subalgebra $\pi
_{\theta}(\mathcal{N})$ in $\mathcal{N}\otimes L^{\infty}(G)$ through
$(\pi_{\theta}(X)\xi)(g):=\theta_{g}^{-1}(X)(\xi(g))$ for $\xi\in
L^{2}(G,\mathfrak{H})$. By restriction on the centre $\mathfrak{Z}%
(\mathcal{N})=\mathcal{A}$ of $\mathcal{N}$ the action $\theta$ of $G$ defines
a W*-dynamical system $\mathfrak{Z}(\mathcal{N})\underset{\beta}%
{\curvearrowleft}G$, which we call a central W*-dynamical system. The
corresponding crossed product $\mathcal{Q}:=\mathfrak{Z}(\mathcal{N}%
)\rtimes_{\beta}G$ can be regarded as a subalgebra of $\mathcal{N}%
\rtimes_{\theta}G$. We recall that a W*-dynamical system $\mathcal{N}%
\underset{\theta}{\curvearrowleft}G$ is \textit{ergodic} if its $\theta$-fixed
point algebra is trivial: $\mathcal{N}^{\theta}=\mathbb{C}1$, and is called
\textit{free} if there exists $X\in\mathcal{N}$ for any non-zero $A\in$
$\mathcal{N}$ and $s\in G$, $s\neq e$, such that $\beta_{s}(X)A\neq AX$.\ When
applied to the abelian algebra $\mathcal{A}$, the latter condition is
equivalent to the requirement that,\ for any compact subset $K\subset
G\setminus\{e\}$ and a non-zero projection $P\in\mathcal{A}$, there exists a
non-zero projection $E\in\mathcal{A}$ such that $E\leq P$\ and $E\beta
_{s}(E)=0$ for any $s\in K $. For an action of an \textit{abelian} group, its
ergodicity automatically implies that it is free.\ The ergodicity and freeness
of the action $\beta$ in the W*-dynamical system $\mathcal{A}\underset{\beta
}{\curvearrowleft}G$ are related with the algebraic properties of the
corresponding crossed product $\mathcal{Q}=\mathcal{A}\rtimes_{\beta}G$ in the
following way:

\begin{proposition}
For an abelian W*-dynamical system $\mathcal{A}\underset{\beta}%
{\curvearrowleft}G$ and the corresponding von Neumann algebra $\mathcal{Q}%
=\mathcal{A}\rtimes_{\beta}G$,

\begin{enumerate}
\item[(i)] the action $\beta$ is free if and only if $\mathcal{A}$ is
maximally abelian in $\mathcal{Q}$: $\mathcal{A}$ $=\mathcal{Q}\cap
\mathcal{A}^{\prime}$;

\item[(ii)] when $\beta$ is free, $\mathcal{Q}$ is a factor if and only if
$\beta$ is ergodic. In this case, the centre of $\mathcal{Q}$ is equal to
$\mathcal{A}^{\beta}$: $\mathfrak{Z}(\mathcal{Q})=\mathcal{A}^{\beta}$.
\end{enumerate}
\end{proposition}

(The proofs of the above proposition and of all the following statements are
omitted here, which will be given in a separate paper \cite{OT06}.)

In view of the close relations between the W*-dynamical systems $\mathcal{N}%
\underset{\theta}{\curvearrowleft}G$ and its central subsystem $\mathfrak{Z}%
(\mathcal{N})\underset{\beta}{\curvearrowleft}G$, the action $\theta$ is said
to be \textit{centrally ergodic} if its restriction $\beta$ is ergodic on
$\mathfrak{Z}(\mathcal{N})$, and \textit{centrally\ free} if$\ \beta$ is free
on $\mathfrak{Z}(\mathcal{N})$. We can verify some commutant relations between
$\mathfrak{Z}(\mathcal{N})$ and $\mathcal{N}$ in $\mathcal{M}$ valid for a
centrally free action $\theta$, which plays essential roles for the analysis
of $\mathcal{M}$:

\begin{proposition}
The following relations hold:
\begin{align*}
\pi_{\beta}(\mathfrak{Z}(\mathcal{N)})  &  =\mathcal{M}\cap\pi_{\theta
}(\mathcal{N})^{\prime},\\
\pi_{\theta}(\mathcal{N})  &  =\mathcal{M}\cap\pi_{\beta}(\mathfrak{Z}%
(\mathcal{N)})^{\prime},\\
\pi_{\theta}(\mathcal{N}^{\theta})  &  =\mathcal{M}\cap\mathcal{Q}^{\prime}.
\end{align*}

\end{proposition}

While $\mathfrak{Z}(\mathcal{N})\cong\mathcal{A}$ is not maximal abelian in
$\mathcal{M}=\mathcal{N}\rtimes_{\theta}G$ owing to the above relations,
$\mathcal{A}$ can be easily extended to a MASA $\mathcal{R}=\mathcal{A}%
\otimes\mathcal{L}$ in $\mathcal{N}$ by tensoring a MASA $\mathcal{L=L}%
^{\prime}$ in\ $B(\mathfrak{H})$. If $\theta$ is centrally free, the
subalgebra $\pi_{\theta}(\mathcal{R})$ is shown to be a MASA in $\mathcal{M}$.

The central ergodicity of $\theta$ is related with the factoriality of the
crossed product:

\begin{corollary}
If $\theta$ is a centrally free action, we have
\[
\mathfrak{Z}(\mathcal{M})=\mathfrak{Z}(\mathcal{Q})=\pi_{\beta}(\mathfrak{Z}%
(\mathcal{N)}^{\beta}).
\]
Therefore the following conditions are equivalent:

\begin{enumerate}
\item[(i)] the action $\theta$ is centrally ergodic;

\item[(ii)] $\mathcal{M}$ is a factor;

\item[(iii)] $\mathcal{Q}$ is a factor.
\end{enumerate}
\end{corollary}

Note that the action $\theta$ is free because $G$ is abelian, and is ergodic
when $\mathcal{M}$ is a factor. $\mathcal{Q}=\mathcal{A}\rtimes_{\beta}G$ is a
factor of finite type only when $G$ is discrete and, otherwise, is properly
infinite. 

\subsection{Modular structure and von Neumann types of $\mathcal{M}$ in terms
of observable data}

To extract the modular data from\ the dynamical system $\mathcal{N}%
\underset{\theta}{\curvearrowleft}G$ necessary for the classification of the
Micro-algebra $\mathcal{M}=\mathcal{N}\rtimes_{\theta}G$, we consider a n.f.s.
weight $\varphi$ of the von Neumann algebra $\mathcal{N}$. Let $(\pi_{\varphi
},U_{\varphi},\mathfrak{H}_{\varphi},J_{\varphi},\mathcal{P}_{\varphi})$ be
the corresponding standard representation of $\mathcal{N}\underset{\theta
}{\curvearrowleft}G$ which consists of the semi-cyclic representation
$(\pi_{\varphi},\mathfrak{H}_{\varphi})$ of $\mathcal{N}$ associated to
$\varphi$, a modular conjugation operator $J_{\varphi}$, a natural positive
cone $\mathcal{P}_{\varphi}$ and the covariant representation $U_{\varphi}$ of
$\theta$ in $\mathfrak{H}_{\varphi}$, $\pi_{\varphi}(\theta_{g}(X))=U_{\varphi
}(g)\pi_{\varphi}(X)U_{\varphi}(g)^{\ast}$. As $\mathcal{N}$ is considered in
$\mathfrak{H}_{\varphi}$, we omit here the symbol $\pi_{\varphi}$ identifying
$\mathcal{N}$ with $\pi_{\varphi}(\mathcal{N})$.

Using the covariant representation $U_{\varphi}$, we construct a left Hilbert
algebra in the representation space $\mathfrak{H}_{\varphi}\otimes
L^{2}(G)=L^{2}(G,\mathfrak{H}_{\varphi})$ of $\mathcal{N}\rtimes_{\theta}G $
\cite{TakII} as follows. First in the space $\mathcal{C}_{c}(G,\mathcal{N}) $
of $\sigma$*-strongly continuous $\mathcal{N}$-valued functions on $G$ with
compact supports, a convolution and an involution are defined for
$X,Y\in\mathcal{C}_{c}(G,\mathcal{N}),$\ $s,t\in G$ by
\begin{align*}
(X\ast Y)(s)  &  =\int_{G}X(t)\theta_{t}\left(  Y(t^{-1}s)\right)  ds,\\
X^{\sharp}(s)  &  =\theta_{s}\left(  X(s^{-1})^{\ast}\right)  .
\end{align*}
With the left and right actions of $A\in\mathcal{N}$ on $X\in C_{c}%
(G,\mathcal{N})$ defined by
\[
(A\cdot X)(s):=AX(s),\qquad(X\cdot A)(s):=X(s)\theta_{s}(A),\text{ \ \ \ for
}s\in G,
\]
$C_{c}(G,\mathcal{N})$ is a bimodule over $\mathcal{N}$ satisfying the
compatibility conditions:
\begin{align*}
A\cdot(X\ast Y)  &  =(A\cdot X)\ast Y,\qquad(X\ast Y)\cdot A=X\ast(Y\cdot
A),\\
(X\cdot A)^{\sharp}  &  =A^{\ast}\cdot X^{\sharp},\qquad\qquad\quad(A\cdot
X)^{\sharp}=X^{\sharp}\cdot A^{\ast},
\end{align*}
for $A\in\mathcal{N}$ and $X,Y\in C_{c}(G,\mathcal{N})$.

Next we denote by $\mathcal{K}_{\varphi}:=Lin\{X\cdot A:X\in C_{c}%
(G,\mathcal{N}),A\in\mathcal{N}\}$, the linear hull of $C_{c}(G,\mathcal{N}%
)\cdot\mathcal{N}$. Since $\mathfrak{n}_{\varphi}:=\{A\in\mathcal{N}$;
$\varphi(A^{\ast}A)\}$ is a left ideal, we have $Y(s)A\in\mathfrak{n}%
_{\varphi}$ for $Y\in\mathcal{C}_{c}(G,\mathcal{N})$, $A\in\mathfrak{n}%
_{\varphi}$, $s\in G$, and hence, $\eta_{\varphi}{\large (}Y(s)A{\large )}%
=Y(s)\eta_{\varphi}(A)$ is meaningful. Accordingly, $\eta_{\varphi}%
{\large (}X(s){\large )}$ makes sense for $X\in\mathcal{K}_{\varphi}$, and we
see that a function $G\ni s\mapsto\eta_{\varphi}{\large (}X(s){\large )}%
\in\mathfrak{H}_{\varphi}$ belongs to $\mathcal{C}_{c}(G,\mathcal{N})$. With a
map $\widetilde{\eta}_{\varphi}$ from $\mathcal{K}_{\varphi}$ to
$L^{2}(G,\mathfrak{H}_{\varphi})$ is defined by
\[
\widetilde{\eta}_{\varphi}(X)(s)=\eta_{\varphi}{\large (}X(s){\large )},\qquad
X\in\mathcal{K}_{\varphi},\text{ \ }s\in G,
\]
$\widetilde{\mathfrak{A}}_{\varphi}:=\widetilde{\eta}_{\varphi}(\mathcal{K}%
_{\varphi}\cap\mathcal{K}_{\varphi}^{\#})$ is a left Hilbert algebra equipped
with the following product and involution:
\begin{align*}
\widetilde{\eta}_{\varphi}(X)\widetilde{\eta}_{\varphi}(Y) &  =\widetilde
{\eta}_{\varphi}(X\ast Y),\\
\widetilde{\eta}_{\varphi}(X)^{\sharp} &  =\widetilde{\eta}_{\varphi}%
(X^{\#}),\qquad\ \ \ X,Y\in\mathcal{K}_{\varphi}\cap\mathcal{K}_{\varphi}%
^{\#}.
\end{align*}
With the definition,
\[
\widetilde{\pi}_{\theta}(X):=\int_{G}X(s)(U_{\varphi}(s)\otimes\lambda_{s})ds,
\]
we see the following relation
\[
\widetilde{\eta}_{\varphi}(X\ast Y)=\widetilde{\pi}_{\theta}(X)\widetilde
{\eta}_{\varphi}(Y),\qquad\ \ \ X,Y\in\mathcal{K}_{\varphi}\cap\mathcal{K}%
_{\varphi}^{\ast},
\]
which shows the equality $\pi_{l}{\large (}\widetilde{\eta}_{\varphi
}(X){\large )}=\widetilde{\pi}_{\theta}(X)$ for the left multiplication
$\pi_{l}$ on $\widetilde{\mathfrak{A}}_{\varphi}$. Therefore $\widetilde{\pi
}_{\theta}$ is a *-representation of $\mathcal{K}_{\varphi}\cap\mathcal{K}%
_{\varphi}^{\ast}$, and $\widetilde{\pi}_{\theta}(\mathcal{K}_{\varphi}%
\cap\mathcal{K}_{\varphi}^{\ast})$ generates the crossed product
$\mathcal{N}\rtimes_{\theta}G$ which is isomorphic with the left von Neumann
algebra $\mathcal{R}_{l}(\widetilde{\mathfrak{A}}_{\varphi})$ of
$\widetilde{\mathfrak{A}}_{\varphi}$. Therefore, the modular structure of the
crossed product $\mathcal{N}\rtimes_{\theta}G$ is determined by the standard
form $(\pi_{\varphi},\mathfrak{H}_{\varphi},J_{\varphi},\mathcal{P}_{\varphi
})$ of $\mathcal{N}$. The modular operator $\widetilde{\Delta}$ and modular
conjugation $\widetilde{J}$ are given by
\begin{align*}
\left(  \widetilde{\Delta}^{it}\xi\right)  (s) &  =\Delta_{\varphi\circ
\theta_{s},\varphi}^{it}\xi(s),\\
\left(  \widetilde{J}\xi\right)  (s) &  =U_{\varphi}(s)J_{\varphi}\xi
(s^{-1}),\qquad\xi\in L^{2}(G,\mathfrak{H}_{\varphi}),\quad s\in G,
\end{align*}
where $\Delta_{\varphi\circ\theta_{s},\varphi}$ is the relative modular
operator from $\varphi$ to $\varphi\circ\theta_{s}$ with which Connes cocycle
derivative $V_{t}=(D(\varphi\circ\theta_{s}):D\varphi)_{t}$ is related
through$\ \Delta_{\varphi\circ\theta_{s},\varphi}^{it}=V_{t}\Delta_{\varphi
}^{it}$ (see \cite{Con}). The dual weight $\widehat{\varphi}$ of
$\mathcal{R}_{l}(\widetilde{\mathfrak{A}}_{\varphi})=\mathcal{N}%
\rtimes_{\theta}G$ is defined by such a n.f.s. weight induced from the left
Hilbert algebra $\widetilde{\mathfrak{A}}_{\varphi}$ as given for
$X\in\mathcal{N}_{+}$ by
\[
\widehat{\varphi}(X)=\left\{
\begin{array}
[c]{c}%
\text{\ }\left\Vert \xi\right\Vert ^{2},\qquad\ \ \ X=\pi_{l}(\xi)^{\ast}%
\pi_{l}(\xi),\quad\xi\in\mathfrak{B},\\
+\infty,\qquad
\ \ \ \ \ \ \ \ \ \ \ \ \ \ \ \ \ \ \ \ \ \ \ \ \ \ \ \ \ \ \ \ \ \ \ \ \ \ \ \ \
\end{array}
\right.
\]
where $\mathfrak{B}$ is the set of left bounded vector in $\widetilde
{\mathfrak{A}}_{\varphi}$. The modular automorphism group $\sigma
\widehat{^{\varphi}}$ of $\widehat{\varphi}$ is given by $\sigma_{t}%
^{\widehat{\varphi}}(X)=\widetilde{\Delta}^{it}X\widetilde{\Delta}^{-it}$ for
$X\in\mathcal{N}\rtimes_{\theta}G$, whose action on the generators
$\pi_{\theta}(\mathcal{N})$, $\lambda(G)$ of $\mathcal{N}\rtimes_{\theta}G$
can be specified explicitly by:
\begin{align*}
\sigma_{t}^{\widehat{\varphi}}{\large (}\pi_{\theta}(X){\large )} &
=\pi_{\theta}(\sigma_{t}^{\varphi}(X)),\qquad\ X\in\mathcal{N},\quad
t\in\mathbb{R},\\
\sigma_{t}^{\widehat{\varphi}}{\large (}\lambda(s){\large )} &  =\lambda
(s)\pi_{\theta}{\large (}(D\varphi\circ\theta_{s}:D\varphi)_{t}{\large )}%
,\qquad s\in G.
\end{align*}

As $\mathcal{N}=\mathcal{A}\otimes B(\mathfrak{H})$ is not finite, its crossed
product $\mathcal{M}$ is not either and we have the following theorem:

\begin{theorem}
For a centrally ergodic W*-dynamical system $(\mathcal{N}\underset{\theta
}{\curvearrowleft}G)$ with its corresponding central W*-dynamical system
$(\mathfrak{Z}(\mathcal{N)}\underset{\beta}{\curvearrowleft}G)$, the factor
type of $\mathcal{M}=\mathcal{N}\rtimes_{\theta}G$ coincides with that of
$\mathcal{Q}=\mathfrak{Z}(\mathcal{N)}\rtimes_{\beta}G$ and we have the
following criteria:

\begin{enumerate}
\item[(i)] $\mathcal{M}$ is of type I if and only if $(\mathfrak{Z}%
(\mathcal{N)}\underset{\beta}{\curvearrowleft}G)$ is isomorphic to the flow on
$L^{\infty}(G)$: $(\mathfrak{Z}(\mathcal{N)}\underset{\beta}{\curvearrowleft
}G)\cong(L^{\infty}(G)\underset{Ad\lambda_{G}}{\curvearrowleft}G)$;

\item[(ii)] $\mathcal{M}$ is of type II if and only if $(\mathfrak{Z}%
(\mathcal{N)}\underset{\beta}{\curvearrowleft}G)$ is not isomorphic to
$(L^{\infty}(G)\underset{Ad\lambda_{G}}{\curvearrowleft}G)$ and $\mathfrak{Z}%
(\mathcal{N)}$ admits a $\beta$-invariant semifinite measure supported by
$\mathfrak{Z}(\mathcal{N)}$;

\item[(iii)] $\mathcal{M}$ is of type III if and only if $\mathfrak{Z}%
(\mathcal{N)}$ admits no $\beta$-invariant semifinite measure with support
$\mathfrak{Z}(\mathcal{N)}$.
\end{enumerate}
\end{theorem}

It is remarkable that the modular structure of $\mathcal{M}$ is completely
determined by the properties of the abelian dynamical system $\mathcal{A}%
\underset{\beta}{\curvearrowleft}G$. In more details in the above type
classification, the spectrum $Spec(\mathcal{A})$ of the centre $\mathcal{A}%
=\mathfrak{Z}(\mathcal{N)=}\mathfrak{Z}(\mathcal{M}\rtimes_{\alpha}%
\mathcal{U})$ of the composite system $\mathcal{M}\rtimes_{\alpha}\mathcal{U}$
plays the crucial role as the classifying space of intrasectorial structure,
in sharp contrast to the quantum-mechanical part $B(L^{2}(\mathcal{U}))$
playing no role. The former is in harmony with the general strategy adopted in
Sec.2 in the sense that intrasectorial analysis reduces to the sector analysis
of the composite system $\mathcal{M}\rtimes_{\alpha}\mathcal{U}$ and the
latter is consistent with the interpretation of it in Sec. 3 as a
(non-commutative) homotopy term. If $\mathcal{M}$ is of type III, we need more
detailed characterization of the modular spectrum $S(\mathcal{M})$ which is
also determined by $\mathcal{A}$ and the action of $\widehat{\mathcal{U}}$ on
it:
\[
S(\mathcal{M})=\bigcap\{Spec(\Delta_{\varphi\circ\theta_{\gamma},\varphi
}):\varphi\in\mathcal{W}_{\mathcal{A}}\},
\]
where $\mathcal{W}_{\mathcal{A}}$ is the set of all normal semi-finite
faithful weights on $\mathcal{A}$ and $\Delta_{\varphi\circ\theta_{\gamma
},\varphi}=\left(  D(\omega\circ\theta_{\gamma}):D\omega\right)  _{t}%
\Delta_{\varphi}$. We recall here the Connes classification of type III von
Neumann algebras \cite{Con}: (1) $\mathcal{M}$ is type III$_{\lambda}$,
($0<\lambda<1$), if and only if $S(\mathcal{M})=\{\lambda^{n}:n\in
\mathbb{Z}\}\cup\{0\}$, (2) $\mathcal{M}$ is type III$_{0}$ if and only if
$S(\mathcal{M})=\{0,1\}$, (3) $\mathcal{M}$ is type III$_{1}$ if and only if
$S(\mathcal{M})=\mathbb{R}_{+}$. \ 

While the type classification does not provide the whole data necessary for
the complete recovery of Micro-algebra without such a uniqueness result as
ensured for the AFD factor of type III$_{1}$, we can draw immediately some
important lessons from it: the above (i) tells us that our starting assumption
on the $\mathcal{U}$-action $\alpha=Ad$ on $\mathcal{M}$\ was too restrictive
to recover $\mathcal{M}$ of non-type I, since it implies that the
corresponding coaction $\theta=\hat{\alpha}$ becomes isomorphic to the flow on
$L^{\infty}(\widehat{\mathcal{U}})$. Recalling the presence of a non-trivial
dynamics inherent to the system $\mathcal{M}$ to be observed, however, we can
easily see that the measurement process described by the coupling $\alpha
_{u}=Ad(u)$ is simply a convenient approximation commonly adopted in most
discussions and that $\alpha$ should \textit{not} be inner in general. To
reconstruct the non-trivial algebra $\mathcal{M}$ of the observed system we
need the data of the \textit{intrinsic dynamics} of the system, which can be
attained by measuring locally the energy-momentum tensor $T_{\mu\nu}$. For
instance, we can approximate the dynamics locally on a subalgebra
$\mathcal{M}=\pi(\mathfrak{A}(\mathcal{O}))^{\prime\prime} $ of local
observables by the modular automorphism group corresponding to a local KMS
state constructed by the Buchholz-Junglas method of heating-up \cite{BuchJung}%
, according to which the above fixed-point algebra $\mathcal{M}^{\alpha
(\mathcal{U)}}$ becomes of type II when $\mathcal{M}$ is of type III. Thus,
starting from $\mathcal{M}\rtimes_{\alpha}\mathcal{U}\simeq\mathcal{A}\otimes
B(\mathfrak{H})$ with $\mathcal{A}$ of type II$_{1}$, we can repeat the
similar analysis to the one in Sec.3. According to the Takesaki duality
\cite{T1}, the crossed product $\mathcal{N}:\mathcal{=M}\rtimes_{\sigma
^{\varphi}}\mathbb{R}$ with respect to the modular automorphism group of a
n.f.s. weight $\varphi$ of $\mathcal{M}$ is a von Neumann algebra of type
II$_{\infty}$ with an n.f.s. trace $\tau$ such that $\tau\circ\theta
_{s}=e^{-s}\tau,s\in\mathbb{R}$ with $\theta$ the action of $\mathbb{R}$ on
$\mathcal{N}$ dual to $\sigma^{\varphi}$, and conversely, $\mathcal{M=N}%
\rtimes_{\theta}\mathbb{R}$ holds. This decomposition of $\mathcal{M}$ is
unique. Furthermore if $\mathcal{M}$ is a factor of type III$_{1}$, then
$\mathcal{N}$ is a factor of type II$_{\infty}$. In such situations, we need
some definitions in relation with a MASA $\mathcal{A}$ of $\mathcal{M}$.
First, the normalizer of $\mathcal{A}$ in $\mathcal{U}(\mathcal{M})$ is
defined by
\[
N_{\mathcal{M}}(\mathcal{A}):=\{u\in\mathcal{U}(\mathcal{M}):u\mathcal{A}%
u^{\ast}=\mathcal{A}\}.
\]
A MASA $\mathcal{A}$ in a factor $\mathcal{M}$ is called \textit{regular} if
$N_{\mathcal{M}}(\mathcal{A})$ generates $\mathcal{M}$, and
\textit{semi-regular} if $N_{\mathcal{M}}(\mathcal{A})$ generates a subfactor
of $\mathcal{M}$. The subalgebra $\mathcal{A}\cap\mathcal{N}$ is also a MASA
of $\mathcal{N}$, for which we can derive the following result from \cite{Pop}:

\begin{proposition}
Let $\mathcal{N=M}\rtimes_{\sigma^{\varphi}}\mathbb{R}$ be a type II$_{\infty
}$ factor von Neumann algebra defined above as a crossed product of a type
III$_{1}$ factor von Neumann algebra $\mathcal{M}$ acting on a separable
Hilbert space $\mathfrak{H}$ with an n.f.s. weight $\varphi$. Then
$\mathcal{N}$ contains a maximal abelian subalgebra $\mathcal{A}$ which is
also maximal abelian in $\mathcal{M}$ and semi-regular in $\mathcal{N}$.
Moreover, if $\mathcal{N}$ is approximately finite dimensional, then
$\mathcal{A}$ can be chosen to be regular in $\mathcal{N}$.
\end{proposition}

In this way, several important steps for the formulation of a measurement
scheme have been achieved in a form applicable to general quantum systems with
\textit{infinite degrees of freedom} as QFT, by removing the restriction on
the choice of MASA inherent to the finite quantum systems and by specifying
the coupling term necessary for constructing measurement processes. To be
fair, however, we note that there remain some unsettled problems, such as the
non-uniqueness of MASA $\mathcal{A}=\mathcal{A}^{\prime}\cap\mathcal{M}$,
which is one of the difficulties caused by the infinite dimensional
non-commutativity. For lack of the uniqueness of MASA the uniqueness of the
above reconstruction of $\mathcal{M}$ is not guaranteed either. In relation to
this, the consistency problem should be taken serious between the
mathematically relevant structures of type III and the \textit{finite
discrete} spectra inevitable at the operational level, which is closely
related to such type of criteria as the nuclearity condition in algebraic QFT
to select the most relevant states and observables. In this connection, it
would be important to re-examine the general meaning of the so-called
\textquotedblleft ambiguity of interpolating fields\textquotedblright\ closely
related to the notion of Borchers classes, relative locality and PCT
invariance. At the end, we remark that the focal point in our consideration
has shifted from states to algebra, from algebra to dynamics, through which
all the basic ingredients constituting a mathematical framework for describing
a physical qunatum system can and/or should be re-examined and re-constructed
in close relations with observational and operational contexts.

\section{Acknowledgments}

Both of the authors would like to express their sincere thanks to Professors
H. Araki, Y. Nakagami, M. Ozawa for their invaluable remarks and comments,
which have been very instructive and helpful. They are also very grateful to
Professors T. Hida, M. Ohya and D. Sternheimer for their interest in the
present project and encouragements. One of the authors (I.O.) was partially
supported by JSPS Grants-in-Aid (No. 15540117).

\end{document}